\documentclass[aps,pre,preprint]{revtex4}
\usepackage{epsfig}
\usepackage{latexsym}
\usepackage{graphicx}

\begin{document}

\title{Finite-Size Scaling for Directed Percolation Models}

\author{Santanu Sinha and S. B. Santra}

\affiliation{Department of Physics, Indian Institute of Technology
Guwahati, Guwahati-781039, Assam, India}

\begin{abstract}
A simple finite-size scaling theory is proposed here for anisotropic
percolation models considering the cluster size distribution function
as generalized homogeneous function of the system size and two
connectivity lengths. The proposed scaling theory has been verified
numerically on two different anisotropic percolation models.
\end{abstract}

\maketitle

\section{Introduction}
Percolation, a model of disorder\cite{perco}, shows a geometrical
phase transition at the percolation threshold characterized by
singularities of cluster related quantities similar to the critical
phenomena or second order phase transition in thermodynamic
systems\cite{statm}. The singularities occurred in phase transitions
are generally described by power laws characterized by well defined
critical exponents. The set of critical exponents and the scaling
relations among them characterize the universality class of a
system. Since most systems are not solvable analytically, numerical
methods are employed in order to investigate the critical behavior. At
the same time, the numerical results are very often limited by the
finite system size. In a finite system, there is rounding and shifting
of critical singularities depending on the ratio of correlation length
$\xi$ to the linear dimension $L$ of the system. In order to obtain
the behavior of the infinite systems, the results of finite systems
are generally extrapolated using finite-size scaling
(FSS)\cite{fss}. However, there are very few FSS theory for
anisotropic models of statistical physics\cite{fss,binder}. FSS theory
is available for isotropic percolation models such as ordinary
percolation\cite{opfss} and spiral percolation (percolation under
rotational constraint)\cite{bose}. Anisotropic percolation clusters
are generated if an external space fixed directional constraint is
applied on the percolation model. Such models are directed percolation
(DP)\cite{hinr} and directed spiral percolation (DSP)\cite{dsp}. In
these models, there are two connectivity lengths $\xi_\parallel$ and
$\xi_\perp$ and they become singular at the percolation threshold with
two different critical exponents. Considering DP as a minimal
stochastic Markovian process represented by Langevin equation, a FSS
theory has been developed by Janssen {\em et al}\cite{jans} below the
upper critical dimension $d_c=4$ and above $d_c$ by L{\"u}beck and
Janssen\cite{lub}. Usually, the anisotropy considered in the FSS
theory is either in the interaction between the constituent
particles\cite{binder} or in the topology of the system (strip like
structure)\cite{fss}. However, in a geometrical model like
percolation, the particles are simply occupied with a probability $p$
in absence of any interaction between them. The models are generally
defined on topologically isotropic systems of size $L\times L$. For
such noninteracting anisotropic models defined on topologically
isotropic systems, a simple phenomenological FSS theory is proposed
here. The proposed FSS theory for anisotropic percolation models is
developed here considering the cluster size distribution function as a
generalized homogeneous function\cite{statm} of occupation probability
$p$ and ratios of connectivity lengths with the system size.

\section{Anisotropic Percolation Models}
There are two well known anisotropic percolation models DP\cite{hinr}
and DSP\cite{dsp}. In DP, a directional field $E$ is present in the
model. The direction of applied $E$ field from upper left to the lower
right corner of the lattice is considered here. As an effect of the
$E$ field, the empty sites to the right and to the bottom of an
occupied site are only eligible for occupation. The eligible sites are
then occupied with probability $p$. Accordingly, the clusters grow in
the diagonal direction along $E$. In the case of DSP, a crossed
rotational field $B$ is also present in addition to the directional
field $E$. In this problem, $E$ field is applied from left to right in
the plane of the lattice and $B$ is applied perpendicular to $E$ and
into the plane of the lattice (viewed from top). Due to $E$ field,
empty site on the right of an occupied site is eligible for occupation
whereas for $B$ field, empty sites in the forward and clockwisely
rotational directions are eligible for occupation. The forward
direction is the direction from which the present site is occupied.
Because of the simultaneous presence of both the $E$ and $B$ fields
crossed to each other, a Hall field appears in the system
perpendicular to both $E$ and $B$. As a result, an effective
directional constraint $E_{\sf eff}$ acts on the system along the left
upper to right lower diagonal of the lattice. The clusters grow along
the effective field $E_{\sf eff}$\cite{dsp}. A cluster is considered
to be a spanning cluster, if either the lateral or the vertical
extension of a cluster becomes equal to the dimension of the lattice
$L$. At the percolation threshold $p_c$, a spanning cluster appears
for the first time in a system. It was found as $p_c \approx 0.705489$
for DP\cite{hinr} and $p_c \approx 0.6550$ for DSP\cite{dsp} on the
square lattice in case of site occupation. For a given finite system
$L$, the cluster properties are calculated generating finite clusters.

\begin{figure}
\centerline{\psfig{file=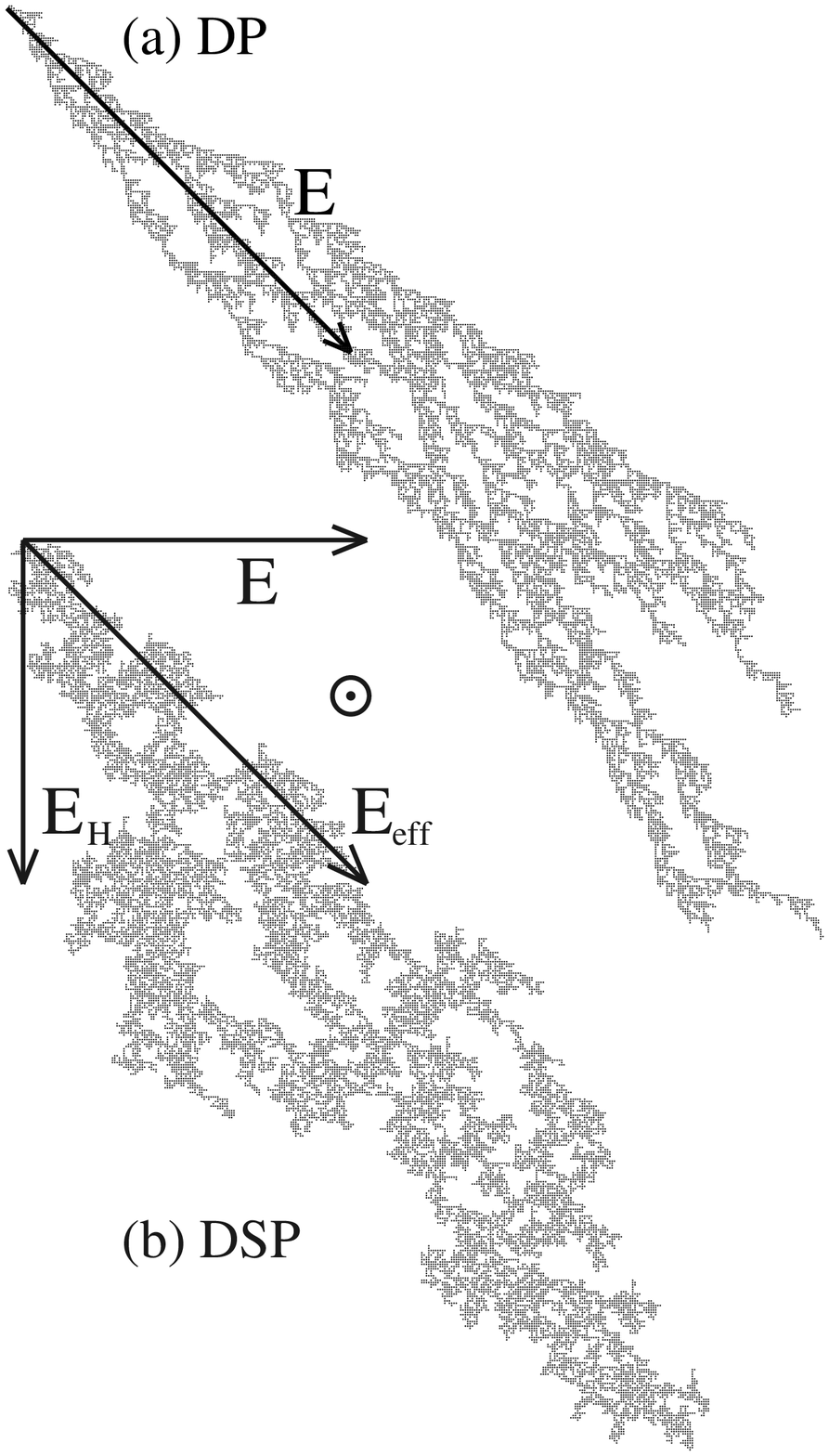,scale=0.55}}
\medskip
\caption{\label{cluster}Typical large clusters of ($a$) directed
  percolation and ($b$) directed spiral percolation generated at
  $p=p_c$ on a square lattice of size $L=256$. Arrows represent the
  directional field $E$ and the encircled dot represents the
  rotational field $B$. }
\end{figure}

Typical large clusters for DP and DSP generated on $256\times 256$
square lattice at their respective percolation thresholds are shown in
Fig.\ref{cluster}. It can be seen that the clusters are anisotropic
and rarefied. In order to characterize the cluster's connectivity
property, two length scales $\xi_\parallel$ and $\xi_\perp$ are
required. $\xi_\parallel$ is the extension along the elongation of the
cluster and $\xi_\perp$ is the extension in the perpendicular
direction of the elongation. Two connectivity lengths $\xi_\parallel$
and $\xi_\perp$ diverge with different critical exponents
$\nu_\parallel$ and $\nu_\perp$ at the $p=p_c$. Though the DSP cluster
grows along an effective directional constraint $E_{\sf eff}$, the
critical properties of the clusters at $p=p_c$ were found different
from that of DP clusters. Accordingly DP and DSP belong to two
different universality classes\cite{dsp}.

\section{FSS Theory}
A system is said to be finite if the system size $L<\xi$, the
connectivity length. In the case of anisotropic percolation models,
there are two connectivity lengths $\xi_\parallel$ and $\xi_\perp$
where $\xi_\parallel$ always greater than $\xi_\perp$. The finiteness
of the system size then can be defined in terms of
$\xi_\parallel$. According to the theory of critical phenomena,
thermodynamic functions become generalized homogeneous functions at
the critical point\cite{statm}. The cluster size distribution function
$P_s(p,L)$ describing the geometrical quantities here in percolation
is then expected to be a generalized homogeneous function at the
percolation threshold $p_c$. A simple phenomenological FSS theory for
these anisotropic percolation models is proposed here assuming
$P_s(p,L)$ as a generalized homogeneous function. In order to develop
FSS theory, the scaling of the order parameter of the percolation
transition $P_\infty$, probability to find a site in a spanning
cluster, with the system size $L$ is considered first. At the end, the
scaling form will be generalized for arbitrary cluster related
quantity $Q$.

In the case of an infinite system, the order parameter $P_\infty$
becomes singular at $p= p_c$ as
\begin{equation}
\label{pinf}
P_\infty\sim (p-p_c)^\beta 
\end{equation}
with a critical exponent $\beta$. In a finite geometry with
topologically symmetric dimension $L\times L$, the critical
singularity of $P_\infty$ for anisotropic percolation clusters depends
not only on $(p-p_c)$ but also on the the ratios $\xi_\parallel/L$ and
$\xi_\perp/L$. The functional dependence of $P_\infty$ on these
parameters is then given by
\begin{equation}
P_\infty(L,p)= {\mathcal F}[(p-p_c),\frac{\xi_\parallel}{L},
\frac{\xi_\perp}{L}].
\label{qfs}
\end{equation}
For an infinite system, at $p=p_c$, $\xi_\parallel$ is infinitely
large and the system properties become independent of the system size
$L$.  Accordingly, two parameters $\xi_\parallel/L$ and $\xi_\perp/L$
in $P_\infty$ given in Eq.\ref{qfs} can be reduced to a single
parameter $\xi_\parallel/\xi_\perp$. Thus, $P_\infty$ in Eq.\ref{qfs}
can be expressed as
\begin{equation}
P_\infty = {\mathcal G}[(p-p_c), \xi_\parallel/\xi_\perp].
\label{pi1}
\end{equation}  
It is now important to know how $\xi_\parallel$ and $\xi_\perp$ scale
with the dimension $L$ of the finite system at the percolation
threshold $p_c$. Two new scaling forms for $\xi_\parallel$ and
$\xi_\perp$ with $L$ are assumed as,
\begin{eqnarray}
\xi_\parallel \approx L^{\theta_\parallel} & \textsf{and} & \xi_\perp
\approx L^{\theta_\perp}
\label{thetdef}
\end{eqnarray}
where $\theta_\parallel$ and $\theta_\perp$ are two new exponents. It
is now possible to define $P_\infty$ in terms of the system size $L$
as
\begin{equation}
P_\infty = F[(p-p_c), L^{\theta_\parallel - \theta_\perp}]. 
\label{pi0}
\end{equation}
Assuming the order parameter $P_\infty$ as a generalized homogeneous
function of $(p-p_c)$ and $L^{\theta_\parallel-\theta_\perp}$, it
is possible to express $P_\infty$ as
\begin{equation}
F[\lambda^a(p-p_c),\lambda^bL^{\theta_\parallel-\theta_\perp}]
  =\lambda P_\infty
\end{equation}
where $a$ and $b$ are arbitrary numbers and $\lambda$ is a parameter.
The above relation is valid for any value of $\lambda$. For $\lambda =
L^{-(\theta_\parallel-\theta_\perp)/b}$, $P_\infty$ takes the form
\begin{equation}
P_\infty = L^{A(\theta_\parallel-\theta_\perp)}
F[(p-p_c)L^{-B(\theta_\parallel-\theta_\perp)}, 1]
\label{pi3}
\end{equation}
where $A=1/b$ and $B=a/b$ are two exponents to be
determined. However, as $L \rightarrow \infty$, the $L$ dependence of
$P_\infty$ will vanish. Therefore, $F[z]$ should go as $z^{A/B}$ in
the limit $L \rightarrow \infty$ when $z =
(p-p_c)L^{-B(\theta_\parallel-\theta_\perp)}$. In that case,
\begin{eqnarray}
P_\infty & \approx & L^{A(\theta_\parallel-\theta_\perp)}
\left(L^{-B(\theta_\parallel-\theta_\perp)}(p-p_c)\right)
^{A/B}\nonumber\\ & \approx & (p-p_c)^{A/B}.
\end{eqnarray}
The order parameter exponent $\beta$ is then given by $\beta = A/B$.

The size of a cluster is given by the number of occupied sites $s$ in
that cluster. For anisotropic clusters, there are two connectivity
lengths which could be measured in terms of radii of gyrations
$R_\parallel$ and $R_\perp$ with respect to two principal axes of the
cluster. The size of the cluster can also be calculated in terms of
area defined by $R_\parallel$ and $R_\perp$. It is expected that the
cluster size should scale as $s \approx R_\parallel R_\perp^{(d_f-1)}$
at $p=p_c$ and it should go as $s \approx R_\parallel R_\perp^{(d-1)}$
above $p_c$ where $d$ is the spatial dimension of the lattice and
$d_f$ is the fractal dimension of the infinite clusters generated on
the same lattice. The percolation probability $P_\infty$ is the ratio
of the number of sites on the infinite cluster to the total number of
sites and can be given as
\begin{equation}
P_\infty = \frac{R_\parallel R_\perp^{(d_f-1)}} {R_\parallel
  R_\perp^{(d-1)}} = R_\perp^{(d_f-d)}.
\label{pidef}
\end{equation}
Assuming $R_\perp \approx \xi_\perp \approx L^{\theta_\perp}$ at
$p_c$, Eq.\ref{pidef} leads to $P_\infty \sim
L^{\theta_\perp(d_f-d)}$. Also at $p=p_c$, the functional form of
$P_\infty$, given in Eq.\ref{pi3}, reduces to $P_\infty \sim
L^{A(\theta_\parallel - \theta_\perp)}$. Therefore, exponent $A$ can
be obtained in terms of the new exponents $\theta_\parallel$ and
$\theta_\perp$ as
\begin{equation}
\label{expa}
A = \frac{\theta_\perp(d_f-d)}{\theta_\parallel-\theta_\perp}.
\end{equation}
Inserting the value of $A$ in Eq.\ref{pi3} at $p=p_c$ it reduces to 
\begin{equation}
P_\infty = L^{\theta_\perp(d_f-d)}F[0] = \xi_\perp^{(d_f-d)}F[0] =
\xi_\parallel^{(d_f-d)\theta_\perp/\theta_\parallel}F[0]
\end{equation}
where $F[0]$ is a constant. For an infinite system, at $p=p_c$, the
connectivity lengths diverge as
\begin{equation}
\label{conl}
\xi_\parallel \sim |p-p_c|^{-\nu_\parallel} \hspace{0.5cm}\and
\hspace{0.5cm} \xi_\perp \sim |p-p_c|^{-\nu_\perp}
\end{equation}
where $\nu_\parallel$ and $\nu_\perp$ are connectivity exponents. The
following scaling relations then can easily be extracted as
\begin{eqnarray}
\label{hsr2}
\beta = \nu_\perp (d-d_f) & \textsf{and} & \beta = \nu_\parallel
(d-d_f)\theta_\perp/\theta_\parallel. 
\end{eqnarray}
The first one of these relations is the hyperscaling
relation\cite{perco} and the second one is a new scaling relation
connecting the exponents $\theta_\perp$ and $\theta_\parallel$. Using
the above scaling relations and eliminating $(d-d_f)$, the values of
the exponents $A$ and $B$ can be obtained as
\begin{eqnarray}
A =-\frac{\theta_\perp\beta}{(\theta_\parallel-\theta_\perp)\nu_\perp}
=-
\frac{\theta_\parallel\beta}{(\theta_\parallel-\theta_\perp)\nu_\parallel}
\nonumber\\ B = -
\frac{\theta_\perp}{(\theta_\parallel-\theta_\perp)\nu_\perp} = -
\frac{\theta_\parallel}{(\theta_\parallel-\theta_\perp)\nu_\parallel}.
\end{eqnarray}
From the expressions of $A$ and $B$, it can be seen that
$\nu_\parallel/\theta_\parallel = \nu_\perp/\theta_\perp$. This is
consistent with the assumption $\xi_\perp\sim
\xi_\parallel^{\theta_\parallel/\theta_\perp}$ in Eq.\ref{thetdef} and
the fact $\xi_\perp\sim \xi_\parallel^{\nu_\parallel/\nu_\perp}$ for
an infinite system. Consequently, one can define the anisotropy
exponent
\begin{equation}
\label{anse}
\theta = \theta_\parallel/\theta_\perp=\nu_\parallel/\nu_\perp
\end{equation}
as suggested in Ref.\cite{willi}. This is interesting to note that
this equation is always valid even if the hyperscaling relations are
not exactly satisfied. It is known that hyperscaling relations are not
satisfied in case of DP\cite{privman}.

Since the values of $A$ and $B$ are now known, the finite-size scaling
form of $P_\infty$ can be given as
\begin{equation}
P_\infty(L,p) = L^{-\beta\theta_\parallel/\nu_\parallel}
F[(p-p_c)L^{\theta_\parallel/\nu_\parallel}] =
L^{-\beta\theta_\perp/\nu_\perp}
F[(p-p_c)L^{\theta_\perp/\nu_\perp}].
\label{pifss}
\end{equation}
It is interesting to note that a finite size scaling relation is
obtained in terms of two unknown exponents $\theta_\parallel$,
$\theta_\perp$ and two known exponents $\nu_\parallel$, $\nu_\perp$.

Now, the finite size scaling form of the order parameter $P_\infty$
can be generalized for any cluster related quantity $Q$. Suppose that
in an infinite system, as $p\rightarrow p_c$ the cluster related
quantity $Q$ scales as
\begin{equation}
\label{qsc}
Q \sim |p-p_c|^{-q}
\end{equation}
where $q$ is a critical exponent. In a finite geometry, it is then
expected that the cluster related quantities in general will obey the
finite size scaling law given by
\begin{equation}
Q(L,p) = L^{q\theta_\parallel/\nu_\parallel}
F[(p-p_c)L^{\theta_\parallel/\nu_\parallel}] =
L^{q\theta_\perp/\nu_\perp}
F[(p-p_c)L^{\theta_\perp/\nu_\perp}].
\label{qfss}
\end{equation}
It should be mentioned here that the finite size scaling obtained here
for the anisotropic cluster is very similar to that of isotropic
clusters. In isotropic FSS, $Q(L,p) = L^{q/\nu} F[(p-p_c)L^{1/\nu}]$
where $\nu$ is the connectivity length exponent\cite{opfss}. In
Eq.\ref{qfss}, $1/\nu$ is replaced by $\theta_\parallel/\nu_\parallel$
or $\theta_\perp/\nu_\perp$ since it is assumed that the connectivity
lengths $\xi_\parallel$ and $\xi_\perp$ scale with the system size $L$
with exponents $\theta_\parallel$ and $\theta_\perp$.  However, it
should be noticed that the FSS relation with respect to $\xi_\perp$,
the transverse length, is new and nontrivial.

The exponents $\theta_\parallel$ and $\theta_\perp$ are measured below
for both DP and DSP clusters and the FSS theory developed here is
applied to the cluster related quantities of both the models.

\section{Verification of FSS Theory}

\begin{figure}
\centerline{\psfig{file=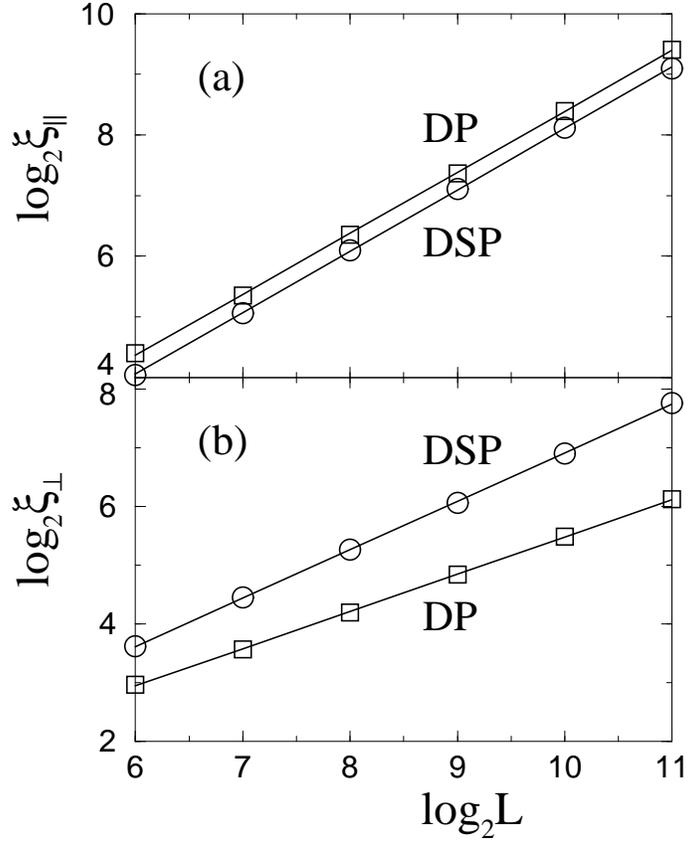,width=0.55\textwidth}}
\medskip  
\caption{\label{theta}Plot of ($a$) $\xi_\parallel$ versus system size
   $L$ and ($b$) $\xi_\perp$ versus $L$ for the DP ($\Box$) and DSP
   ($\bigcirc$) clusters at their respective percolation thresholds.
   From the slopes, values of $\theta_\parallel$ and $\theta_\perp$
   are obtained as $\theta_\parallel=1.01 \pm 0.01$ and
   $\theta_\perp=0.64 \pm 0.01$ for DP clusters. For DSP, the values
   are $\theta_\parallel=1.01 \pm 0.01$ and $\theta_\perp=0.83 \pm
   0.01$ respectively.}
\end{figure}

In order to verify the proposed FSS theory, simulations are performed
on the square lattice of sizes $L=128$ to $2048$ in multiple of
$2$. Average has been taken over $5\times 10^4$ large finite
clusters. First, the exponents $\theta_\parallel$ and $\theta_\perp$
which describes the dependence of connectivity lengths with the system
size $L$ are determined. Since, clusters are grown following single
cluster growth algorithm, the connectivity lengths are given by
$\xi_\parallel^2=2\sum'_sR^2_\parallel sP_s(p,L)/\sum'_ssP_s(p,L)$ and
$\xi_\perp^2=2\sum'_sR^2_\perp sP_s(p,L)/\sum'_ssP_s(p,L)$ where
$R_\parallel$ and $R_\perp$ are radii of gyration with respect to two
principal axes of the cluster. $R_\parallel$ and $R_\perp$ are
estimated from the eigenvalues of the moment of inertia tensor, a
$2\times 2$ matrix here.  The cluster size distribution function
$P_s(p,L)$ is defined as $N_s/N_{tot}$ where $N_s$ is the number of
$s$-sited clusters out of $N_{tot}$ clusters generated on a given
system size. $\xi_\parallel$ and $\xi_\perp$ are measured for various
system sizes $L$ at $p=p_c$ for both DP and DSP clusters and plotted
against the system size $L$ in Fig.\ref{theta}($a$) and
Fig.\ref{theta}($b$) respectively. The squares represent the data of
DP and the circles represent that of DSP. There are two things to
notice. First, the exponent $\theta_\parallel$ is found $\approx 1$
for both the models. It is expected. Because, for the given field
configuration the clusters are elongated along the diagonal of the
lattice and $\xi_\parallel$, extension along the elongation of the
cluster, then should be $\approx \sqrt{2}L$ for large
clusters. Second, the exponent $\theta_\perp$ is found different for
DP and DSP: $\theta_\perp \approx 0.64\pm 0.01$ for DP and
$\theta_\perp \approx 0.83\pm 0.01$ DSP. Assuming
$\theta_\parallel=1$, one should have $\theta_\perp =
\nu_\perp/\nu_\parallel$. Since for DP, $\nu_\perp = 1.0972\pm 0.0006$
and $\nu_\parallel = 1.7334 \pm 0.001$ the expected value of
$\theta_\perp$ is $\approx 0.633$. Similarly for DSP, the expected
value of $\theta_\perp$ is $\approx 0.84$ since $\nu_\perp = 1.12\pm
0.03$ and $\nu_\parallel = 1.33 \pm 0.01$. The measured values are
close to the expected values for both DP and DSP. This implies that
the system properties scale with the system size $L$ at $p=p_c$ in the
similar manner as it behaves with $p$ around $p_c$ for a given large
system size. It can also be noticed that the magnitude of $\xi_\perp$
is different for DP and DSP. $\xi_\perp$ is larger for the DSP
clusters in comparison to that of DP clusters whereas $\xi_\parallel$
of DSP clusters is slightly smaller than that of DP
clusters. Consequently the DSP clusters are less anisotropic than DP
clusters.

\begin{figure}
\centerline{\hfill\psfig{file=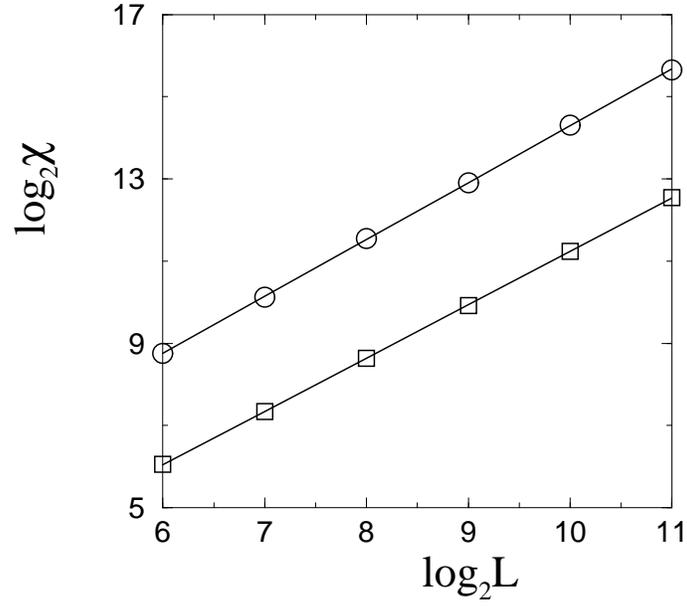,width=0.55\textwidth}\hfill}
\medskip  
\caption{\label{kipc}Plot of average cluster size $\chi$ versus system
  size $L$ for the DP ($\Box$) and DSP ($\bigcirc$) clusters at
  $p_c$. From the slopes, the ratio of
  $\gamma\theta_\parallel/\nu_\parallel$ or
  $\gamma\theta_\perp/\nu_\perp$ is obtained as $1.31 \pm 0.01$ and
  $1.38 \pm 0.01$ for DP and DSP respectively.}
\end{figure}

Next, the average cluster size $\chi$ is measured for different system
size $L$ at $p=p_c$. In single cluster growth algorithm, the average
cluster size is defined as $\chi= \sum_s' sP_s(p,L)$, where $P_s(p,L)$
is the cluster size distribution function. In an infinite system,
$\chi$ diverges as: $\chi\sim |p-p_c|^{-\gamma}$, $\gamma$ is a critical
exponent. According to the FSS theory, it should behave as
$\chi(L)\sim L^{\gamma\theta_\parallel/\nu_\parallel}$ or $\chi(L)\sim
L^{\gamma\theta_\perp/\nu_\perp}$ at $p=p_c$. In Fig.\ref{kipc}, the
average cluster size $\chi$ is plotted against the system size $L$ for
both DP and DSP. The cluster size $\chi$ follows a power law with the
system size $L$. The obtained slopes are $1.31\pm 0.01$ for DP and
$1.38 \pm 0.01$ for DSP. The expected values of the ratio of the
exponents are $\gamma/\nu_\parallel \approx 1.31$ and
$\gamma\theta_\perp/\nu_\perp \approx 1.33$ for DP where $\gamma =
2.2772 \pm 0.0003$. For DSP $\gamma = 1.85 \pm 0.01$ and the expected
ratios of the exponents are: $\gamma/\nu_\parallel \approx 1.39$ and
$\gamma\theta_\perp/\nu_\perp \approx 1.37$. It can be seen that the
measured values are in agreement with that of the expected values
within error bars. It confirms that the cluster properties follow the
proposed finite size scaling theory. It is not surprising that the
scaling theory works with $\xi_\parallel$. However, the proposed
theory works with $\xi_\perp$, the transverse length. This is a new
result.

\begin{figure}
\centerline{\hfill\psfig{file=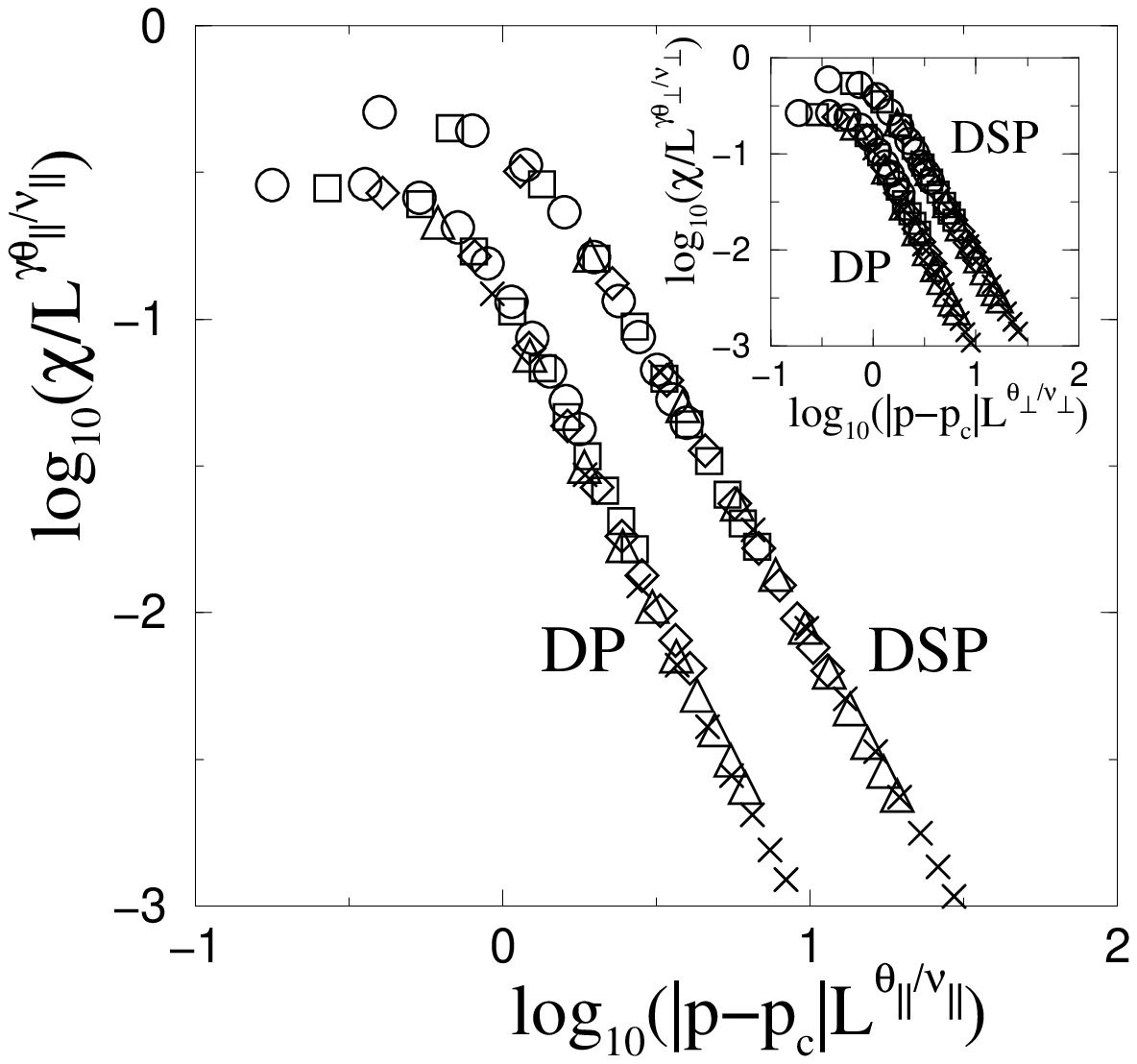,width=0.55\textwidth}
  \hfill}
\medskip  
\caption{\label{scln}Plot of scaled average cluster size $\chi(p,
L)/L^{\gamma\theta_\parallel/\nu_\parallel}$ versus scaled variable
$|p-p_c|L^{\theta_\parallel/\nu_\parallel}$ for DP and DSP clusters
for different values of $L$ and $p$. The data plotted correspond to
different system sizes of $L=128$ ($\bigcirc$), $256$ ($\Box$), $512$
($\Diamond$), $1024$ ($\triangle$) and $2048$ ($\times$). The
$|p-p_c|$ values here are $0.01$ to $0.10$ in the interval of
$0.01$. In the inset, the data collapse is shown plotting $\chi(p,
L)/L^{\gamma\theta_\perp/\nu_\perp}$ versus
$|p-p_c|L^{\theta_\perp/\nu_\perp}$.}
\end{figure}

Finally, the FSS function form has been verified. The average cluster
size is given as $\chi(L,p) = L^{\gamma\theta_\parallel/\nu_\parallel}
F[(p-p_c)L^{\theta_\parallel/\nu_\parallel}] =
L^{\gamma\theta_\perp/\nu_\perp} F[(p-p_c)L^{\theta_\perp/\nu_\perp}]$
for different system size $L$. In Fig.\ref{scln}, the scaled average
size $\chi/L^{\gamma\theta_\parallel/\nu_\parallel}$ is plotted
against the scaled variable
$z=(p-p_c)L^{\theta_\parallel/\nu_\parallel}$. A reasonable data
collapsed is obtained for both DP and DSP. In the inset of
Fig.\ref{scln}, the data collapse is shown by plotting
$\chi/L^{\gamma\theta_\perp/\nu_\perp}$ versus
$z=(p-p_c)L^{\theta_\perp/\nu_\perp}$. It can be seen that the tail of
the scaling function $F(z)$ shows a power law behavior in both the
models with two different scaling exponents, approximately $2.23$ for
DP and $1.86$ for DSP. Once again it confirms that DP and DSP follow
anisotropic finite size scaling and belong to two different
universality classes. Note that, these exponents are close to the
respective cluster size critical exponents $\gamma$ for infinite
systems as expected.

\section{Conclusion}
A finite size scaling theory is proposed here for anisotropic
percolation models like DP and DSP. In this theory the cluster size
distribution is assumed to be a generalized homogeneous function of
the $(p-p_c)$ and the ratios of the connectivity lengths
$\xi_\parallel$ and $\xi_\perp$ to the system size $L$. Reducing the
functional dependence from three variables to two variables $(p-p_c)$
and $\xi_\parallel/\xi_\perp$ and assuming a new scaling form of
$\xi_\parallel$ and $\xi_\perp$ with the system size $L$, a finite size
scaling form of the cluster properties are obtained. Following a
single cluster growth Monte Carlo algorithm, clusters are generated at
the percolation threshold for DP and DSP varying the system size
$L$. The numerical simulation confirms the proposed scaling relations
as well as the scaling function form. It could be considered as a
simplest possible anisotropic finite size scaling theory for directed
percolation models.

\vspace{0.5cm}
\noindent
{\bf Acknowledgment:} SS thanks CSIR, India for financial support.

\newpage

\begin{thebibliography}{18}

\bibitem{perco} A. Bunde and S. Havlin, in {\em Fractals and Disordered
Systems}, edited by A. Bunde and S. Havlin (Springer-Verlag, Berlin,
1991); K. Christensen and N. R. Moloney, {\em Complexity and
Criticality}, (World Scientific, London, 2005).

\bibitem{statm} H. E. Stanley, {\em Introduction to Phase Transitions
and Critical Phenomena}, (Oxford University Press, New York, 1987);
J. M. Yeomans, {\em Statistical Mechanics of Phase Transitions},
(Oxford University Press, New York, 1994).

\bibitem{fss} M. N. Barber, in {\em Phase Transitions and Critical
Phenomena}, Vol. 8, edited by C. Domb and J. L. Lebowitz (Academic
Press, New York, 1984); J. L. Cardy, {\em Finite-size Scaling}, edited
by J. L. Cardy (North Holland, Amsterdam, 1988).

\bibitem{binder} K. Binder and J. S. Wang, J. Stat. Phys {\bf 55},
87 (1989); A. M. Szpilka and V. Privman, Phys. Rev. B {\bf 28},
6613 (1983).

\bibitem{opfss} D. Stauffer and A. Aharony, {\em Introduction to
Percolation Theory}, 2nd edition, (Taylor and Francis, London, 1994).

\bibitem{bose} S. B. Santra and I. Bose, J. Phys. A {\bf 24}, 2367
  (1991) and references there in.

\bibitem{hinr} H. Hinrichsen, Adv. Phys {\bf 49}, 815 (2000) and
references there in.

\bibitem{dsp} S. B. Santra, Eur. Phys. J. B {\bf 33}, 75 (2003);
S. Sinha and S. B. Santra, Eur. Phys. J. B. {\bf 39}, 513 (2004);
S. Sinha and S. B. Santra, Int. J. Mod. Phys. C {\bf 16}, 1251 (2005).

\bibitem{jans} H. K. Janssen, B. Schaub and B. Schmittmann,
Z. Phys. B: Condens. Matter {\bf 71}, 377 (1988).

\bibitem{lub} S. L\"{u}beck and H. K. Janssen, Phys. Rev. E {\bf
72}, 016119 (2005).


\bibitem{essam} J. W. Essam, K. De'Bell, J. Adler and F. M. Bhatti,
Phys. Rev. B {\bf 33}, 1982 (1986); J. W. Essam, A. J. Guttmann and
K. De'Bell, J. Phys. A: Math. Gen. {\bf 21}, 3815 (1988).

\bibitem{willi} W. Kinzel and J. M. Yeomans, J. Phys. A:
Math. Gen. {\bf 14}, L163 (1981); J. K. Williams and N. D. Mackenzie,
J. Phys. A: Math. Gen. {\bf 17}, 3343 (1984).

\bibitem{privman} M. Henkel and V. Privman, Phys. Rev. Lett. {\bf 65},
1777 (1990).

\end{thebibliography}
\end{document}